# Symmetry breaking and structure instability in ultra-thin 2H-TaS$_2$ across charge density wave transition

Divya Rawat, Aksa Thomas, and Ajay Soni*

*School of Physical Sciences, Indian Institute of Technology Mandi, Mandi-175005, Himachal Pradesh, India*

Corresponding Author Email: *ajay@iitmandi.ac.in

## ABSTRACT

Ultra-thin 2D materials have shown complete paradigm shift of understanding of physical and electronic properties because of confinement effects, symmetry breaking and novel phenomena at nanoscale. Bulk 2H-TaS$_2$ undergoes an incommensurate charge density wave (I-CDW) transition temperature, $T_{I\text{-}CDW}$ ~ 76 K, however, onset of CDW in atomically thin layers is not clear. We explored the evidence of CDW instability in exfoliated atomically thin 2H-TaS$_2$ using low temperature Raman spectroscopy. We have emphasized on CDW associated modes, M$_1$ ~ 125 cm$^{-1}$ ($E_1$), M$_2$ ~ 158 cm$^{-1}$ ($A_1$), and M$_3$ ~ 334 cm$^{-1}$ ($E_2$), with thickness ~ 3 nm (one unit-cell). The asymmetric (Fano) line shape of M$_2$ suggests evidence of strong electron-phonon coupling, which mainly drives the CDW instability. Our observations provide key evidence that the CDW can persists even in one-unit cell with a $T_{I\text{-}CDW}$ well above ~ 200 K, which is higher than bulk 2H-TaS$_2$.

Atomically thin two dimensional (2D) transition metal dichalcogenides (TMDCs)[1] have gained significant attentions due to their wide range of electronic, optical, chemical and thermal properties and potential applications field effect transistors[2], optoelectronics[3], photo detectors, energy harvesting and quantum sensing and metrology.[4] Among these 2D materials, the ultra-thin layers of TiSe$_2$,[5] NbSe$_2$,[6] TaSe$_2$,[7] VSe$_2$[8], and TaS$_2$[9] have exotic physical properties due to symmetry breaking with reduced dimensionality and thus showcase the emerging quantum correlated phenomena coupled with the atomic arrangements.[10-14] The quantum size of these material strongly affects the electron correlations, for instance, tunable band gap, charge density wave (CDW), superconductivity, and metal-insulator transitions, which are important to understand the many body physics of electron-electron and electron-phonon interactions,[15] and improved electronic transport.[16] In general, CDW is a periodic modulation of the electronic charge density accompanying with lattice distortion in the material, which is associated with the formation of superlattice and new type of electronic band structure because of the breaking of

*Divya Rawat,* Aksa Thomas and Ajay Soni,
IIT Mandi

the lattice symmetry in the crystal.[17,18] The energy imbalance of CDW is connected with the opening of an electronic band gap within the first Brillouin zone involving both electronic and lattice ordering of the materials. Various theoretical and experimental studies like inelastic neutron and X-ray scattering[19], time resolved optical pump-probe spectroscopy[20] and Raman spectroscopy[21] shows that CDW can be rooted from fermi surface nesting and electron-phonon (*e-ph*) interactions.[17,21-23] CDW can be classified based on the periodicity of the distorted lattice (CDW Wave vector, $\vec{q}_{CDW}$), as commensurate (C-CDW), nearly commensurate (NC-CDW) and incommensurate (I-CDW).[24] Here, the C-CDW has $\vec{q}_{CDW}$ as an integer multiple whereas I-CDW has non-integer multiple of reciprocal lattice vector of the undistorted lattice.[25]

Among layered materials, bulk 2H-TaS$_2$ is metallic in nature and exhibits I-CDW at transition temperature ($T_{I\text{-}CDW}$) between 75-80 K and undergoes superconducting transition at $T_C \sim 0.8$ K.[26] Similar to 2H polymorphs of TMDCs, crystalline unit cell of 2H-TaS$_2$ contains two tri-atomic layers (*S-Ta-S*) separated from each other with weak van der Waals (vdWs) gap.[27] Monolayer of TaS$_2$ is mainly consisting of three atomic planes, where *Ta* is sandwiched between two *S* layers. Previous studies of 1T polymorph of TaS$_2$ have demonstrated the effect of reduced dimensionality on the CDW, however a very few studies available for CDW instability in exfoliated 2H-TaS$_2$. In this regards, the origin mechanism and $T_{I\text{-}CDW}$ of the CDW in ultra-thin layer of 2H-TaS$_2$ is not clear. A recent study revealed that CDW phase cannot exist below a critical thickness (*d*) of ~ 3 nm,[28] however, another report declares the existence of CDW in monolayer of 2H-TaS$_2$[29]. Hence, the CDW instability in thin layers of 2H-TaS$_2$ is highly debatable research, which needs to be explored in detail. The CDW transition temperature ($T_{I\text{-}CDW}$) can be tuned by doping, reduced dimensionality[12], electrostatic gating[14] or strain engineering[30,31] which is already reported for TMDCs like NbSe$_2$,[32] VSe$_2$,[33] TaSe$_2$,[34] TiSe$_2$[35] and 1T-TaS$_2$.[26] Fundamentally, the CDW state in 2H-TaS$_2$ is stabilized by accompanying a periodic lattice distortion majorly through *e-ph* interactions, therefore, the investigation of the phonons is very important for CDW transition in atomically thin flakes.[8] To address this discrepancy, there are various experiments such as charge transport,[36] transmission electron microscopy[28] and inelastic X-ray and neutron scattering[22], have been used so far. Considering the involvements of *e-ph* interaction in CDW, Raman spectroscopy is an easy but a very strong tool to probe the dynamics of phonon instability in thin exfoliated layer of 2H-TaS$_2$.[8]

*Divya Rawat,* *Aksa Thomas and Ajay Soni,*
IIT Mandi

In this work we report on the thickness (*d*) dependent phonon anomaly and structural instability in ultra-thin 2H-TaS$_2$ CDW material. The atomically thin flakes are mechanically exfoliated on SiO$_2$/Si substrate from a single crystal of 2H-TaS$_2$, which is grown by chemical vapor transport technique.[8] The experimental detail regarding Raman, AFM, and transport can be found in the Supporting Information (SI). The $T_{I\text{-}CDW}$ for bulk 2H-TaS$_2$ is estimated using low temperature electrical transport (Figure S1a) and heat capacity measurements (Figure S1b) in SI. Thus, the bulk TaS$_2$ has an expected I-CDW below $T_{I\text{-}CDW}$ ~ 76 K. The bulk crystal and thin flakes are characterized by low temperature Raman and polarization spectroscopy to explore the phonon responses during the CDW instability in the exfoliated thin layer of 2H-TaS$_2$. Thickness, *d,* of the exfoliated flakes have been estimated from the atomic force microscopy (AFM) and corresponding optical images are shown in Figure 1. The high *d* (~ 100 nm) in Figure 1 (a) has been considered as the bulk crystal. Atomically thin layers of the mechanically exfoliated 2H-TaS$_2$ have been first identified by optical microscopy, represented in Figure 1b-d, followed by *d* identifications using topographic images by AFM (insets of Figure 1e-g). The corresponding AFM height profiles have been shown for ~ 10 nm (Figure 1e), ~ 7 nm (Figure 1f), ~ 5 nm and ~ 3 nm as marked in Figure 1g. We have also analyzed the larger flakes and shown in SI, for instance, ~ 27 nm (Figure S2a), ~ 16 nm (Figure S2b).

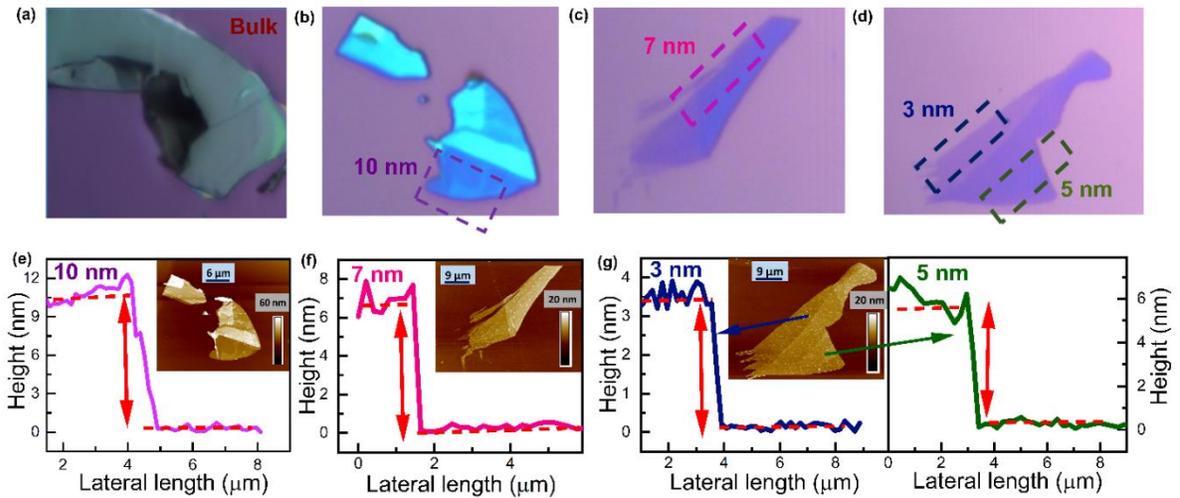

**Figure 1.** Optical images of (a) 2H-TaS$_2$ single crystal and mechanically exfoliated thin flakes (b) ~ 10 nm (c) ~ 7 nm (d) ~ 5 nm and ~ 3 nm on Si/SiO$_2$ substrate. Height profiles of (e) ~ 10 nm (f) ~ 7 nm (g) ~ 5 nm and ~ 3 nm thin flakes and corresponding topographic AFM images are shown in respective insets.

*Divya Rawat*, Aksa Thomas and Ajay Soni,
IIT Mandi

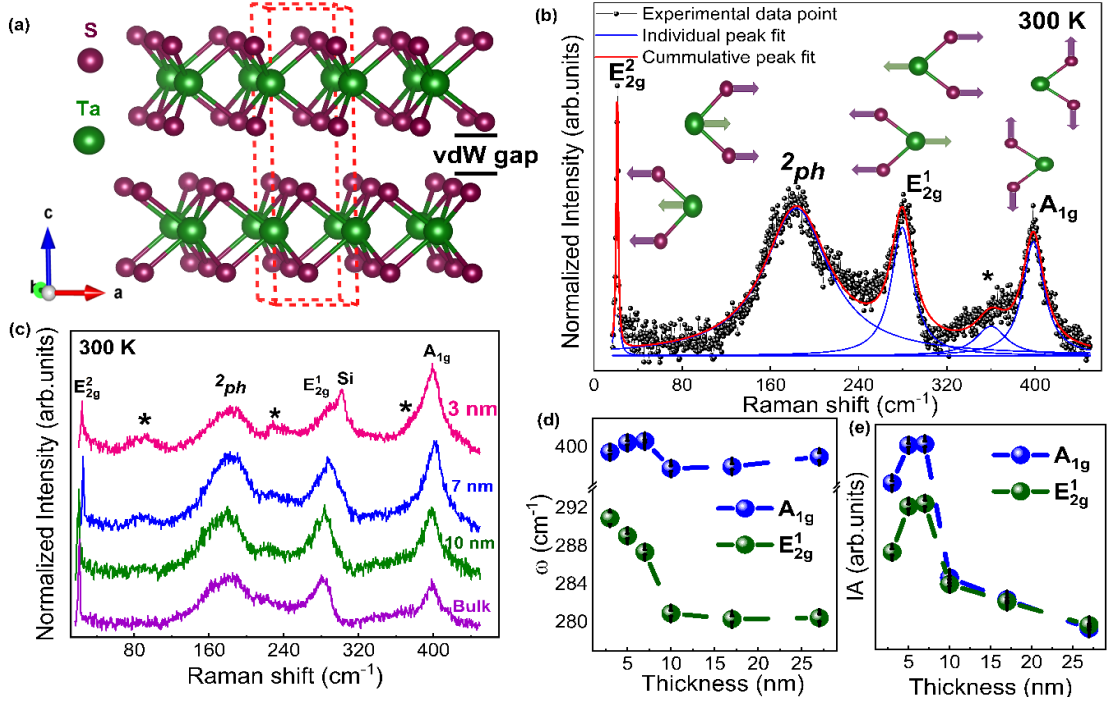

**Figure 2.** (a) Crystal structure of layered 2H-TaS$_2$, where Ta and S atoms are shown with green and magenta color, respectively, and box represents the unit cell. (b) Normalized room temperature Raman spectra of bulk (inset shows the schematic representation of $E_{2g}^2$, $E_{2g}^1$, and $A_{1g}$ Raman modes), Thickness dependent (c) Raman spectra, (d) $\omega$, and (e) *IA* of $E_{2g}^1$, and $A_{1g}$ Raman modes.

The 2H-TaS$_2$ crystallizes in hexagonal phase, wherein a unit cell is composed of two tri-atomic layers (*S-Ta-S*) stacked with weak (vdWs gap (Figure 2a). The tri-atomic layer (called monolayer) is made of covalent bonded *S-Ta-S* in the trigonal prismatic structure, where *Ta* atomic layer is sandwiched between layers of *S* atoms.[27] Thus, primitive cell of 2H-TaS$_2$ has three atoms. Correspondingly, there are nine vibrational modes at the center of the Brillouin zone, three of which are acoustic and six are optical modes. According to the group theory analysis, these optical modes can be represented as irreducible representation; $\Gamma = A_{1g} + E_{1g} + 2E_{2g} + A_{2u} + E_{1u}$.[29, 37] Here, $A_{1g}$, $E_{1g}$, and $E_{2g}$ are Raman active while $A_{2u}$, and $E_{1u}$ are infra-red active modes. Figure 2b has the normalized Raman spectra of the bulk 2H-TaS$_2$ showing various modes at ~ 21 cm$^{-1}$ ($E_{2g}^2$), ~180 cm$^{-1}$ ($2_{ph}$), ~ 236 cm$^{-1}$ ($E_{2g}^1$), and ~ 400 cm$^{-1}$ ($A_{1g}$), and symmetries are assigned in consistent with earlier reports.[21, 36] The schematics of the corresponding atomic displacements of Raman-active modes are represented in inset of Figure 2b. The $A_{1g}$ symmetry has atomic displacements along the out of plane while the $E_{2g}$ has the

*Divya Rawat, Aksa Thomas* and *Ajay Soni,*
IIT Mandi

displacements along the in-plane direction of the flakes. The mode at ~180 cm$^{-1}$ is a characteristic mode of CDW materials and defined as two phonon mode ($2_{ph}$), which is originated from the second order scattering process of two acoustic phonons.[21, 38] Generally, $2_{ph}$ mode is a second order scattering process of two acoustic phonons having frequency approximately twice of Kohn anomaly near CDW wavevector ($\vec{q}_{CDW} = 2/3 \Gamma M$). The mode at ~ 360 cm$^{-1}$, is a arising from the background.[21] A strong dimensionality effect is observed in the Raman spectra of flakes with varying *d*, at 300 K, (Figure 2c). For the ~ 3 nm flakes, the broad modes at ~ 88 cm$^{-1}$, ~ 228 cm$^{-1}$, 380 cm$^{-1}$ are related to the substrate (Si/SiO$_2$).[21] The contribution from the substrate is ubiquitous for atomically thin layers on the substrate and is also observed in Raman spectra of monolayers of TMDCs such as MoS$_2$[39], WS$_2$[40], WSe$_2$[41], and TaSe$_2$[42]. A systematic variation of phonon frequency ($\omega$) and integerated area (*IA*) of the $E_{2g}^1$ and $A_{1g}$ with *d* are summarized in Figure 2d and 2e. A clear *d* dependence on the Raman modes is observed, where the the $E_{2g}^1$ mode softens (red shifts) with increasing *d* (Figure 2d). The softening with *d* is because of the enhancement of the dielectric screening of the long-range coulomb interactions in bulk.[43, 44] On the other hand the $A_{1g}$ mode hardens (blue shifts) for layer *d* up to ~ 7 nm and then drops abruptly for ~ 10 nm and followed by hardening at higher *d*. The abrupt softening of $A_{1g}$ mode in the ~ 10 nm has been observed in earlier reports.[44] The hardening of $A_{1g}$ mode is understood with the classical harmonic oscillators coupled through vdWs forces as the effective restoring force is increasing with addition of layers.[45] Interestingly, the *IA* of both $E_{2g}^1$ and $A_{1g}$ modes (Figure 2e), decreases with the increase in the *d* (except for ~ 3 nm), that indicates the lower cross-section area in thick flakes.[39] The anomalous behavior of *IA* for ~ 3 nm is because of the involvements of substrate contributions in estimating the *IA*.

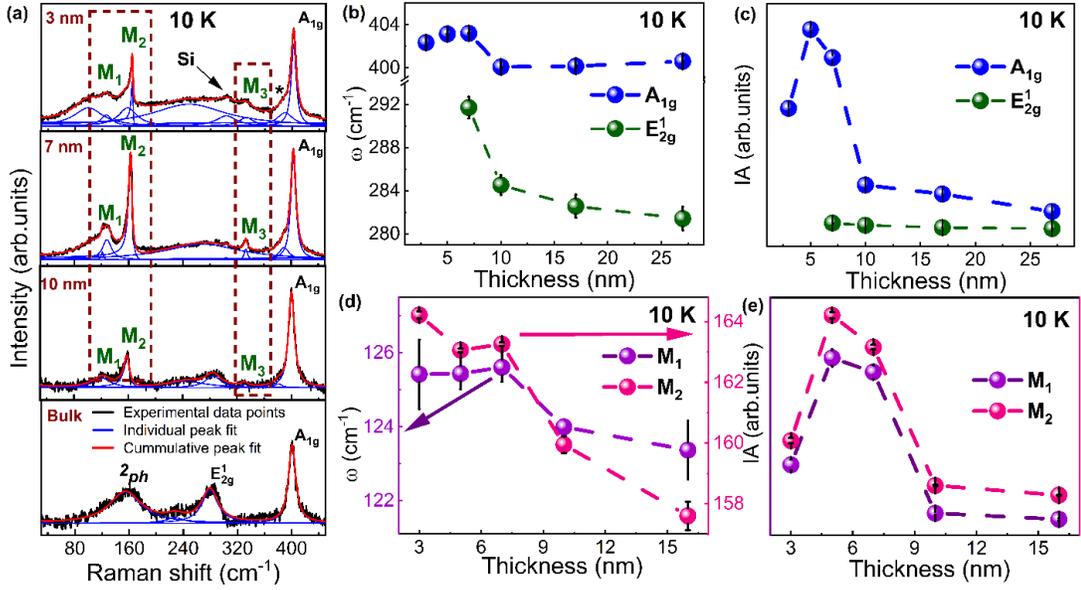

**Figure 3.** (a) Raman spectra at 10 K for bulk, ~ 10 nm, ~ 7 nm, and ~ 3 nm flakes. Thickness dependence $\omega$, and *IA* (b-c) of $E^1_{2g}$, $A_{1g}$, (d-e) $M_1$, and $M_2$ Raman modes, respectively.

Figure 3a shows the Raman spectra, at 10 K, where the bulk crystal has expected $2_{ph}$, $E_{2g}$ and $A_{1g}$ symmetry modes, while the exfoliated flakes shows new Raman modes $M_1$ ~ 125 cm$^{-1}$, $M_2$ ~ 158 cm$^{-1}$, $M_3$ ~ 334 cm$^{-1}$, arising due to dimensionality effect. To the best of our understanding these modes are observed because of CDW transition but not discussed in details, so far. For bulk crystal, the intensity of the $A_{1g}$ modes for spectra at 10 K have increased in compared with spectra at 300 K (Figure 2b). Apart from the $E^1_{2g}$, $A_{1g}$ and *2ph* modes, no signatures of new modes ($M_1$, $M_2$, and $M_3$) observed in the spectra of bulk sample (T$_{CDW}$ ~ 76 K). Thus, $M_1$, $M_2$ and $M_3$ modes are characteristic of thin flakes ($\leq$ 16 nm) and are associated with the lower dimensionality, symmetry breaking, and structural instabilities associated with CDW. The detailed analysis of the new modes is done in following texts. Similar to the *d* dependence at 300 K, both the $E^1_{2g}$ and $A_{1g}$ modes have shown the softening and hardening along with *IA* at 10 K as well (Figure 3b and c), which shows this nature is intrinsic to the flakes. For the one-unit cell, (~ 3 nm), the $E^1_{2g}$ is very broad and the data could not be shown in Figure 3b and 3c. Symmetry of the $M_1$, $M_2$ and $M_3$ modes is confirmed by the polarized Raman spectroscopy (Figure S3).[8, 46] The $M_2$ mode is present in parallel polarization, whereas $M_1$ and $M_3$ are present for both parallel as well as perpendicular polarization. Thus, $M_2$ belong to *A* symmetry while $M_1$ and $M_3$ are of *E* symmetry.[46] The modes $M_1$ and $M_2$ exist between the frequency range of 120-160 cm$^{-1}$ and can associated with mechanism of $2_{ph}$ mode which is observed for bulk crystal. Remarkably, $M_1$ and $M_2$ exhibit distinct

*Divya Rawat,* Aksa Thomas and Ajay Soni,
IIT Mandi

symmetries but show a significant softening with the increase in *d* of the flakes (Figure 3d). The *ω* of $M_1$ is constant below ~ 7 nm and softens for thick flakes, which follows nature of $E^1_{2g}$. However, in the case of $M_2$, which is of *A*-symmetry, the softening with *d* is anomalous and surprising.

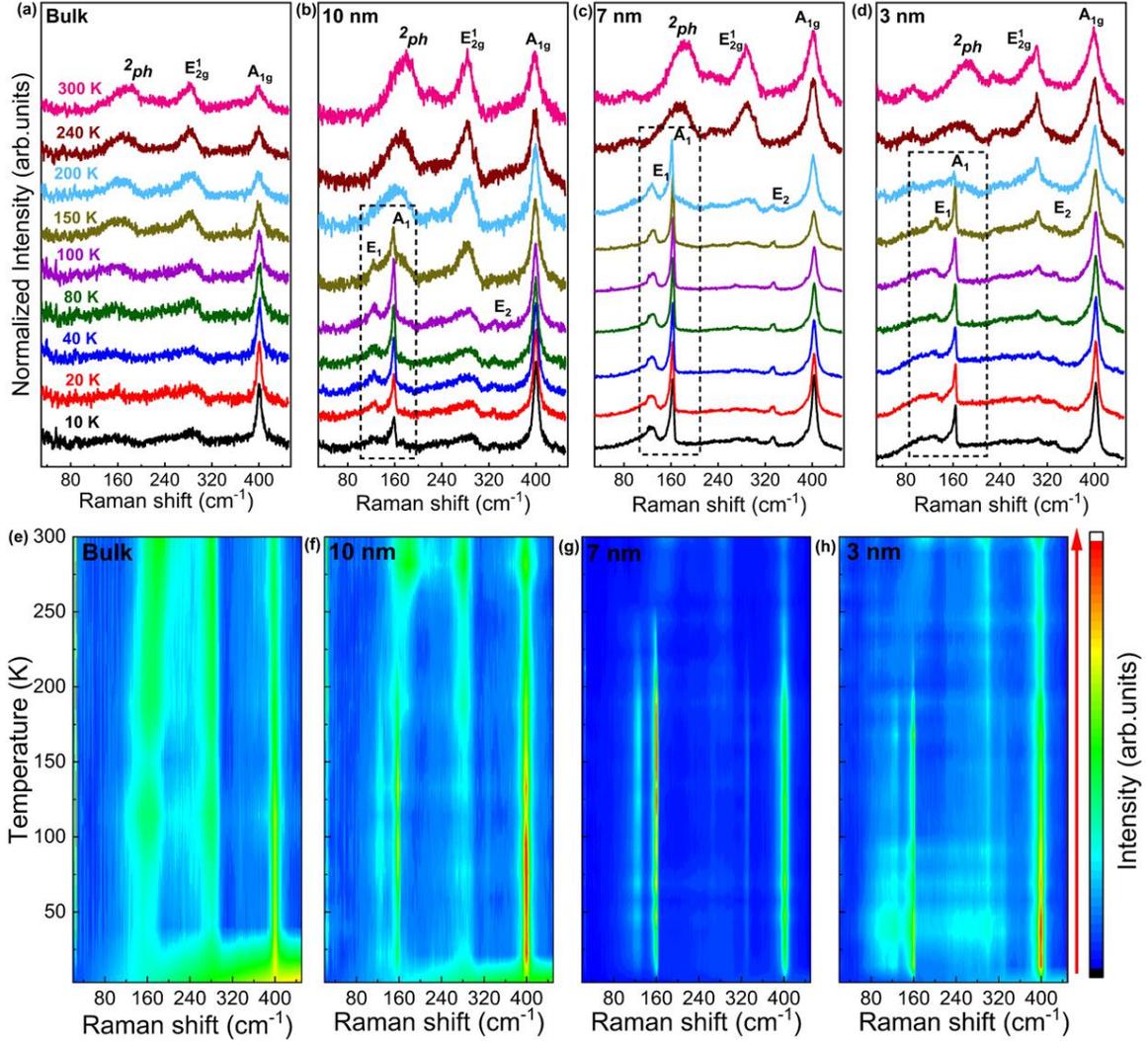

**Figure 4.** Temperature dependent Raman spectra of (a) bulk (b) ~ 10 nm (c) ~ 7 nm (d) ~ 3 nm. Here, rectangle higlights the Raman modes asociated with I-CDW transition. The corresponding intensity color map is shown for (e) bulk, (f) ~ 10 nm (g) ~ 7 nm and (h) ~ 3 nm flakes.

The line shape analysis of $M_2$ show an asymmetry towards low energy indicating the existence of Fano resonance (Figure S4), which is also observed in the graphene[47], $Bi_2Se_3$[45] and nanowires[48]. Fano resonance is a quantum interference effect which depends primarily on the *e-ph* coupling strength of the material.[45, 49] Thus, the anomalous response of $M_2$ and the asymmetric line shape provides significant

*Divya Rawat,* Aksa Thomas and Ajay Soni,
IIT Mandi

evidence of the involved *e-ph* coupling in thin flakes of 2H-TaS$_2$.[21] For 2H-TaS$_2$, the *e-ph* coupling is the major cause of CDW. Similar to the *d* dependence for *IA* of $E_{2g}^1$ and $A_{1g}$ at 300K (Figure 2e) and 10K (Figure 3c), the *IA* of both M$_1$ and M$_2$ also decreases with increasing *d* (Figure 3e). Recently, *Han-Chun Wu et.al* has observed new Raman mode ~ 155 cm$^{-1}$ in monolayer of 2H-TaS$_2$, which is known to be degenerate CDW breathing and wriggling mode.[29] In bulk 2H-TaS$_2$, CDW has been well understood using transport, angle resolved photoemission spectroscopy, density functional theory.[21] To explore the cause and effect of CDW in thin layers and to elucidate the observation of new CDW modes, a detailed temperature dependent studies of ultra-thin 2H-TaS$_2$ have been performed (Figure 4). For bulk and ~ 27 nm thick flake (Figure S5a), the M$_1$, M$_2$, and M$_3$ have not observed even in the low temperature range, while for flakes with *d* below ~ 16 nm (Figure S5b), these modes are observed clearly. The presence of M$_1$, M$_2$, and M$_3$ indicates the existence of structural instability, which appears at temperature, T$_{I-CDW}$, where the I-CDW sets up. For bulk crystal, only CDW collective modes are observed below 80 K,[36] however the structural instability and CDW gap ($\Delta_{CDW}$) has also been reported at ~ 240 K using angle-resolved photoemission spectroscopy and density functional theory analysis.[21] For resistance measurement on monolayer of 2H-TaS$_2$, the reported *T$_{I-CDW}$* is ~ 93 K, which is quite high as compared to bulk 2H-TaS$_2$ i.e. ~ 76 K.[29] Hence, the appearance of the new modes in thin layer can be inferred due to the structural instability or formation of the CDW even at higher temperatures in few-layered than relative to the bulk of 2H-TaS$_2$. The increased CDW transition temperature in thin layer is due to the reduced dimensionality, which enhances *e-ph* coupling and has been observed in other 2D CDWs as well. [50] [5]From the appearance of the M$_1$, M$_2$ and M$_3$ modes, the *T$_{I-CDW}$* of the flakes has been estimated. Corresponding temperature dependent color map of Raman intensity has also been shown for bulk (Figure 4e), ~ 10 nm (Figure 4f), ~ 7 nm (Figure 4g), ~ 3 nm (Figure 4h). For all samples, the intensity of the modes increases at low temperatures. In case of ~ 3 nm, intensity map little noisy because of the substrate effects. Evidence of *e-ph* coupling and commensurability of CDW can be strongly identified by the variation of *ω*, *fwhm*, and *IA* of new modes, M$_1$ and M$_2$.

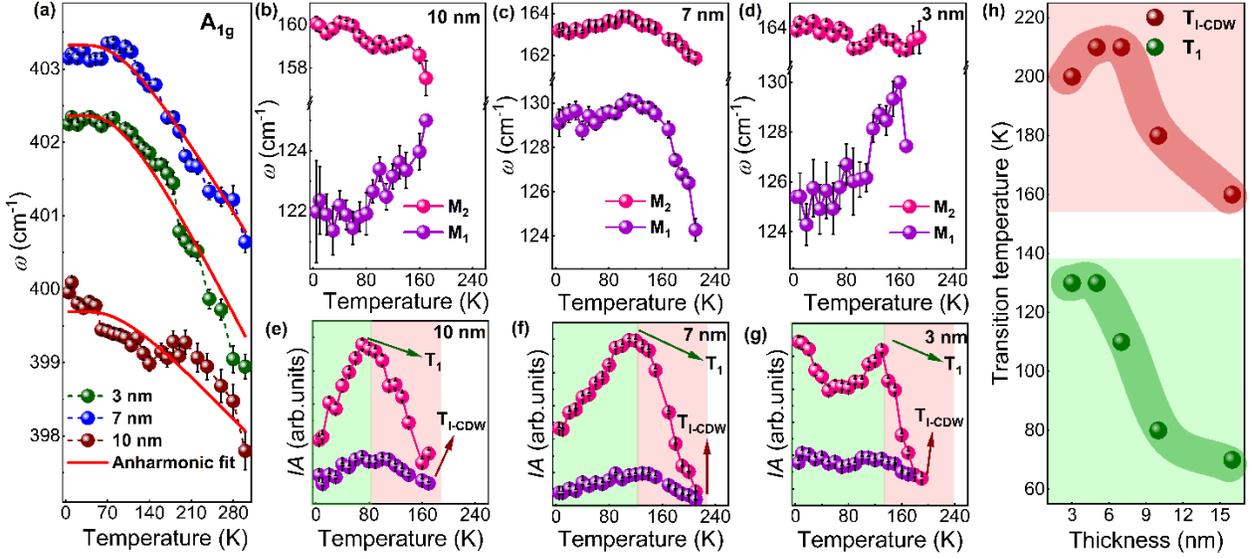

**Figure 5.** Temperature dependence $\omega$ of (a) $A_{1g}$, and of $M_1$ and $M_2$ for (b) ~ 10 nm, (c) ~ 7 nm, and (d) ~ 3nm thin layer. Temperature dependence *IA* (e-f-g), for ~ 10 nm, ~ 7 nm, and ~ 3 nm, respectively. (h) Thickness dependence $T_{I-CDW}$ and $T_1$ of thin layer 2H-TaS$_2$. Solid red line is the fitted data with anharmonic model.

Figure 5 shows the temperature dependence of the ω, and *IA* of CDW modes, $M_1$ and $M_2$, for ~ 10 nm, ~ 7 nm and ~ 3 nm thin flakes. In general the Raman modes softens at higher temperatures because of thermal and volume expansion with the temperatures.[8, 46, 51, 52] The anharmonic behavior of temperature dependence of $A_{1g}$ ~ 400 cm$^{-1}$ mode is shown in Figure 5a and the detailed analysis of ω, *fwhm*, and *IA* of $A_{1g}$ mode is shown in Figure S6. On the other hand, the ω and *IA*, (*fwhm* in Figure S7), of CDW modes ($M_1$ and $M_2$) show anomalous behavior with rise in temperature (Figure. 5b-g). The anomalous behavior in the Raman spectra is a signature of the structural instabilities associated with the CDW in the flakes below $T_{CDW}$ and is related with the formation of CDW superlattices and strong *e-ph* coupling of new CDW modes. Noticeably, the *IA* for all exfoliated flakes has strong thickness dependence and reaches maxima at certain temperature (ordering at $T_1$) followed by a decrease at lower temperatures. This behavior is indication of interaction of density of electronic states with phonons and is related to the structural ordering at low temperature, which is also observed in VSe$_2$ by *Pandey. et.al.*[8] The temperature dependence of the ω, *fwhm*, and *IA* of CDW modes, $M_1$ and $M_2$, for ~ 16 nm, ~ 5 nm thick and thin flakes, has also been shown in SI (Figure. S8). Based on the temperature dependence of the *IA* of the $M_1$ and $M_2$ (Figure 5e-g and Figure S8e-f), two regions have been defined, (i) onset of the I-CDW ($T_{I-CDW}$, green color) and (ii) onset of ordered state ($T_1$, pink color). The *d* dependence of $T_{I-CDW}$ and $T_1$, has been summarized in Figure 5h, indicates that both transition temperature decreases with the increase in the d of the exfoliated flakes. The observation of anamolous behavior of CDW modes ($M_1$, $M_2$, and $M_3$) provide the substantial evidence of *e-ph* coupling, which is suggesting that the CDW is strongly affected by the nanoscale dimensionality of atomically thin 2H-TaS$_2$.

**Conclusion**

In summary the confinement effects and symmetry breaking at nanoscale shows a higher $T_{I-CDW}$ ~ 200 K and presence of CDW modes, $M_1$ ~ 125 cm$^{-1}$ ($E_1$), $M_2$ ~ 158 cm$^{-1}$ ($A_1$), and $M_3$ ~ 334 cm$^{-1}$ ($E_2$), in one unit cell of exfoliated 2H-TaS$_2$. The symmetries and nature of the CDW associated modes is understood from low-temperature and polarized Raman spectroscopy. The evidence of strong electron-phonon coupling has been understood from the asymmetric (Fano) line shape of $M_2$, which may be giving the CDW instabilities. We observed that the $T_{I-CDW}$ has inverse relation with the *d* of the flakes, which suggest existence of CDW instability above $T_{I-CDW}$ of bulk (~ 76 K) 2H-TaS$_2$.

**ASSOCIATED CONTENT**

Supporting Information is available free of charge. It includes the details of synthesis, characterization, Raman measurements, AFM imaging, and additional results and discussions complementary to the main texts.

**ACKNOWLEDGEMENT:** We would like to acknowledge IIT Mandi for the research facility.

*Divya Rawat,* Aksa Thomas and Ajay Soni,
IIT Mandi

Supplementary Information for

# Symmetry breaking and structure instability in ultra-thin 2H-TaS$_2$ across charge density wave transition

Divya Rawat, Aksa Thomas, and Ajay Soni*

*School of Physical Sciences, Indian Institute of Technology Mandi, Mandi-175005, Himachal Pradesh, India*

Corresponding Author Email: *[ajay@iitmandi.ac.in](ajay@iitmandi.ac.in)*

1. **Growth and characterization of single-crystal 2H-TaS$_2$**

Single crystal of 2H-TaS$_2$ was grown by chemical vapor transport technique using iodine as a transporting agent.[8, 53] Thin flakes of crystals were mechanically exfoliated and transferred to SiO$_2$/Si substrate after piranha cleaning. The optical images were captured using the Olympus microscope attached to the Raman spectrometer. The thickness identification of the exfoliated flakes was done by atomic force microscopy (AFM) using Bruker Dimension icon AFM. The estimated thickness of flakes is lying within an error bar of ~ 1 nm. Low temperature resistance and heat capacity measurements were performed by the Quantum Design make physical properties measurement system. Raman spectroscopy was performed using Jobin Vyon LabRAM HR Evolution Raman spectrometer with Czerny-turner grating (1800 gr/mm) and Peltier cooled CCD detector in a back-scattering configuration. A 532 nm laser was used and 50x long working distance objective was used for focusing laser on the sample. Ultra-low frequency filter was used to access low-frequency Raman modes. To control the polarization state, a λ/2 half-waveplate and an analyzer was used in front of the objective lens and spectrometer respectively to select the desired polarization component of the incident and scattered light. Temperature-dependent Raman measurement was performed using a Montana instrument make closed-cycle cryostation. All the Raman spectra were fitted by Lorentzian and Fano model to evaluate full width at half maximum (*fwhm*), phonon frequency (*ω*), and integrated area (*IA*) of optically active Raman modes.

2. **Resistance and Heat-capacity measurements of bulk 2H-TaS$_2$.**

Evidence of the charge density wave (CDW) in bulk 2H-TaS$_2$ has been observed through resistance (*R*) and heat capacity (*C$_p$*) measurement. The broad hump in *R* curve (Figure S1a), indicates the CDW phase transition and the *T$_{I-CDW}$* has been estimated from the *dR/dT* (inset. Figure S1a). The sharp peak in the *C$_p$* data (Figure S1b), also strongly supports the evidence of CDW at ~ 76 K.

3. **Optical and AFM imaging with height profile of exfoliated thick flakes (~ 27 nm and ~ 16 nm)**

Thick films of 2H-TaS$_2$ having thickness ~ 27 nm, and ~ 16 nm have also been obtained through mechanical exfoliation. Optical and AFM images with corresponding height profile have been represented in Figure S2.

4. **Polarized Raman spectroscopy of ~ 7 nm thin layer of 2H-TaS$_2$**

*Divya Rawat, Aksa Thomas and Ajay Soni,*
IIT Mandi

Low temperature Raman spectra at 10 K is intense in case of ~ 7 nm (Figure 2a in manuscript). Therefore, further ~ 7 nm thin flake is chosen to perform the polarization Raman spectroscopy for the identification of the symmetry of new Raman modes ($M_1$, $M_2$, and $M_3$) in thin layer of 2H-TaS$_2$. Here, the laser is incident normal to the *ab* plane and polarization of incident laser is rotated (θ) from 0° to 180° using λ/2 half wave-plate. In polarized Raman spectroscopy, two different configurations have been defined, parallel (*XX*) and perpendicular (*XY*) configuration.[8] In *XX* configuration (θ = 0°) the incident and the scattered light are polarized along the same direction, while for the *XY*, they are orthogonal to each other. Among all new Raman modes, $M_2$ is absent in *XY* configuration, while $M_1$ and $M_3$ are present (Figure S3), indicates that $M_1$ and $M_3$ are of *E*- and $M_2$ is of A-symmetry Raman modes.[46]

### 5. Analysis of asymmetric (Fano) line shape of $M_2$ at 10 K.

The line-shape of $M_2$ has asymmetric shape towards lower energy showing the presence of an interesting Fano resonance, that can provide vital information about the interaction of the electronic continuum with a discrete phonon state, electron-electron interaction and electron-plasmon interaction.[45] The asymmetric line shape for all thin flakes ~ 10 nm (Figure S4a), ~ 7 nm, (Figure S4b) ~ 3 nm (Figure S4c) (~ 10 nm, ~ 7 nm, and ~ 3 nm) was fitted using Breit Wigner Fano (BWF) function, which can be defined by the mathematical expression, $F(\omega) = I_0 \frac{(q+\varepsilon)^2}{(1+\varepsilon^2)}$ , where $\varepsilon = (\omega-\omega_0)/\Gamma$ and $1/q$ is defined as asymmetric parameter.[49] The $1/q$ is commonly interpreted as a measure of the electron-phonon (*e-ph*) coupling strength and it sign indicate the location of charge continuum, whether near valence band or conduction band. Existence of the Fano line-shape, itself provide the significant evidence of *el-ph* coupling[45], which can be interpreted as the interference of a continuum of electronic excitations with the phonon line and plays an important role behind the existence of the new modes.

### 6. Temperature dependent Raman spectra of ~ 27 nm, ~ 16 nm, and ~ 5 nm.

To indicate the evolution of new modes ($M_1$, $M_2$, $M_3$) in thick flakes, temperature dependent Raman spectroscopy have been performed on other exfoliated flakes also (~ 27 nm, ~ 16 nm, and ~ 5 nm), represented in Figure S5. The existence of new modes have been observed for ~ 16 nm thick layer (Figure S5b) also, but absent in case of ~ 27 nm (Figure S5a). From the temperature dependent Raman spectra, we can accomplish that new modes are present only in thin layer ≤ 16 nm. The region of temperature, where the new modes exist have been shown by the dashed rectangle.

### 7. Variation of *fwhm* and *IA* of $A_{1g}$ Raman mode for all thin flakes.

The temperature dependence of ω of $A_{1g}$ mode for thin layer (~ 10 nm, ~ 7 nm, and ~ 3 nm), represented in Figure 5a (Main manuscript) has been fitted with the anharmonic decay model using the equation: $\omega(T) = \omega_o - \frac{A}{e^{\frac{\hbar\omega_o}{2k_BT}}-1}$; where $\omega_o$ is the phonon frequency at 0 K and *A* is the anharmonic constant.[51]

Experimental data points are quite well fitted with the anharmonic model (Figure 5a), indicates the purely anharmonic contribution in usual phonon, which mainly arise because of thermal and volume expansion with the temperature.[8, 51, 52] The temperature dependence of *fwhm* (Figure S6a), *IA* (Figure S6b) also indicates the normal anharmonic behavior because of the thermal and volume expansion.[51]

**8. Variation of *fwhm* of $M_1$ and $M_2$ for ~ 10 nm, ~ 7 nm, and ~ 3 nm thin flakes.**

The temperature dependence of *fwhm* of $M_1$ and $M_2$ for ~ 10 nm (Figure S7a), ~ 7nm (Figure S7b), and ~ 3 nm (Figure S7c), shown the anomalous behavior with respect to normal phonon modes, which indicates the existence of *e-ph* coupling.

**9. Variation of $\omega$, *fwhm*, and *IA* of $M_1$ and $M_2$ for ~ 16 nm, ~ 5 nm thin flakes.**

Figure S8 shows the temperature dependence of the ω, fwhm, and *IA* of CDW modes, $M_1$ and $M_2$, for ~ 16 nm, and ~ 5 nm thin flakes. The $T_{I\text{-}CDW}$ has been estimated from the appearance of CDW modes and $T_1$ has been estimated from the temperature dependence of *IA* which show anomalous behavior below certain temperature. Decrease in the *IA* below certain temperature, indicates the decrease in the electronic density that will correspond to the existence of *e-ph* coupling and commensurability of the formed CDW superlattice.[8]

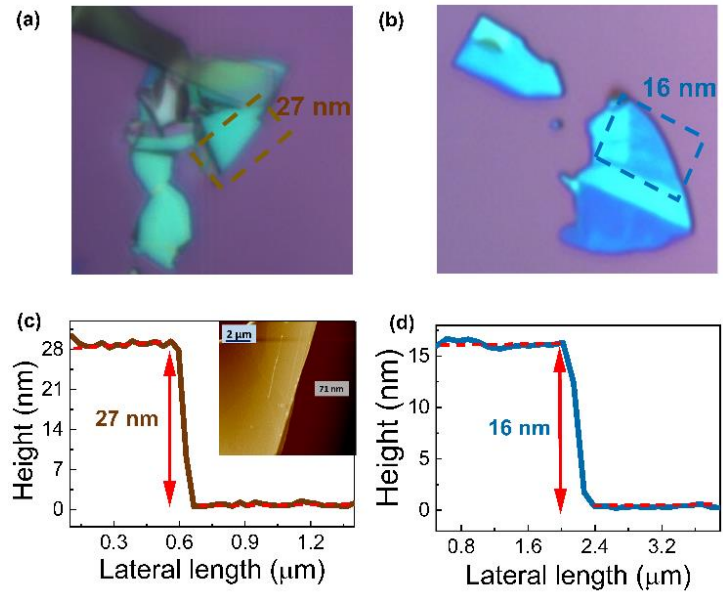

**Figure S2.** Optical images of mechanically exfoliated thin flakes (a) ~ 27 nm (b) ~ 16 nm on Si/SiO$_2$ substrate. The height profiles of (c) ~ 27 nm (d) ~ 16 nm thin flakes and corresponding topographic AFM images are shown in respective insets.

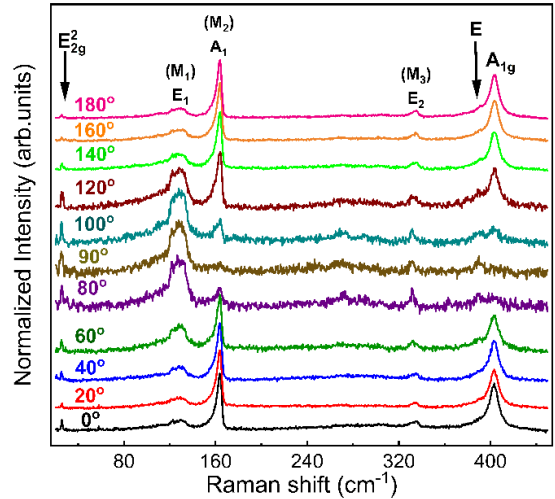

**Figure S3.** Polarized Raman spectra of ~ 7 nm thin layer with the variation of polarization of incident light from 0° to 180° in backscattering geometry.

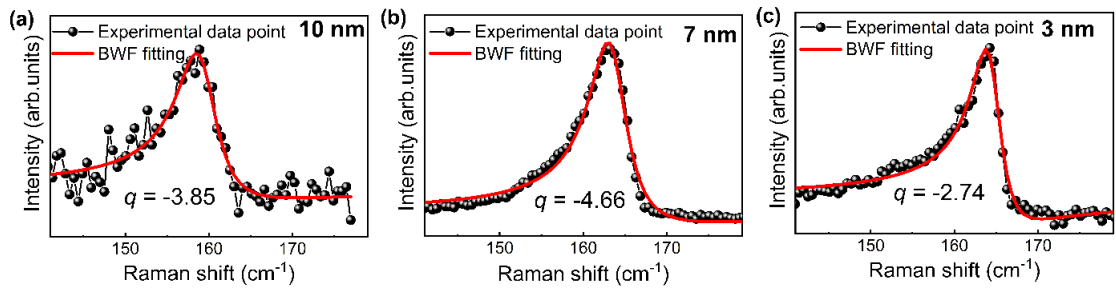

**Figure S4.** Thickness dependence Fano line shape of $M_2$ mode for (a) ~ 10 nm (b) ~ 7 nm, and (c) ~ 3 nm.

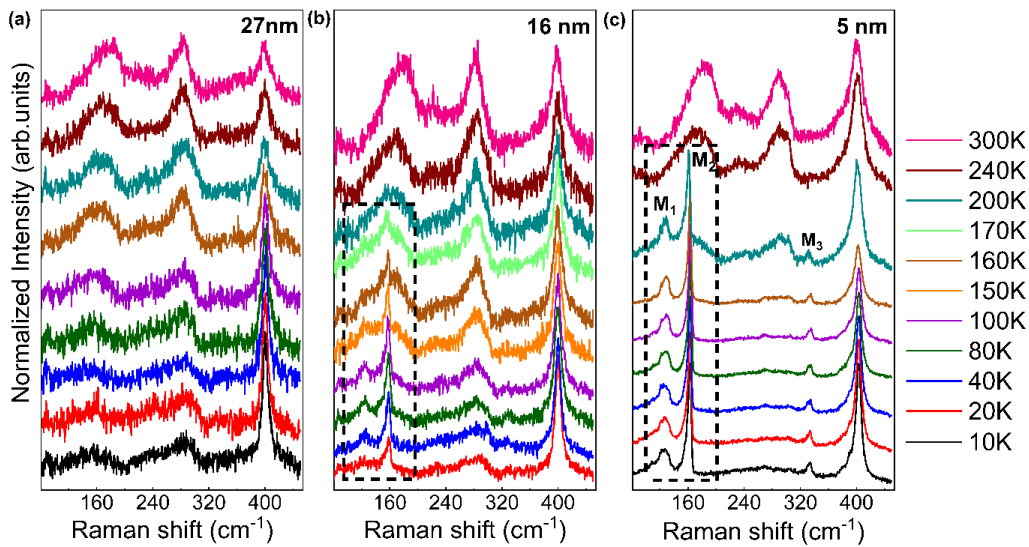

**Figure S5.** Temperature dependent Raman spectra of (a) ~ 27 nm, (b) ~ 16 nm and (c) ~ 5 nm thin layer.

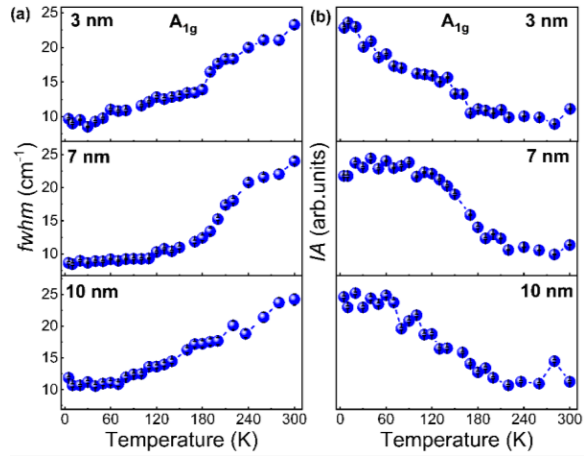

**Figure S6.** Temperature dependent of (a) *fwhm*, and (b) *IA* of $A_{1g}$ ~ 400 cm$^{-1}$ mode.

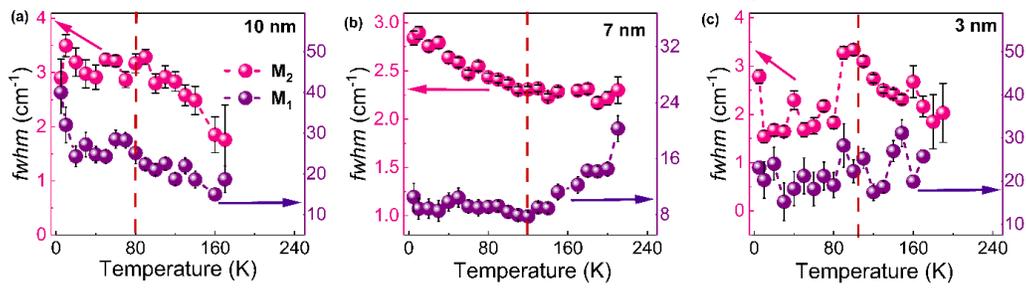

**Figure S7.** Temperature dependence of *fwhm* for (a) ~ 10 nm, (b) ~ 7 nm, and (c) ~ 3 nm.

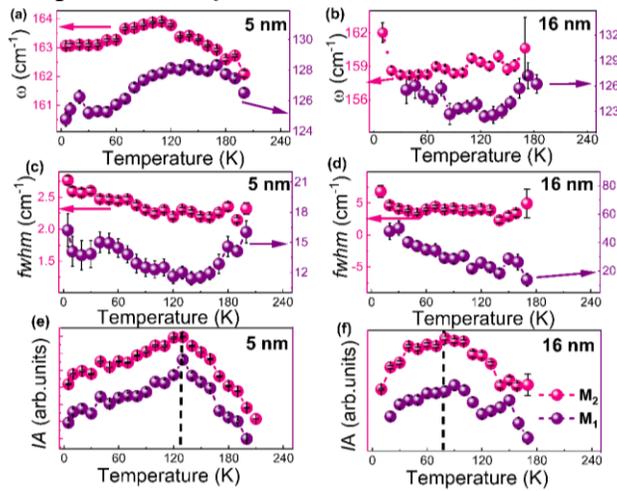

**Figure S8.** Temperature dependence of *ω* (a-b), *fwhm* (c-d), and *IA* (e-f) for ~ 5 nm, and ~ 16 nm, respectively.